\documentclass[a4paper,11pt]{article}
cm \voffset = -2truecm

\usepackage{graphicx}
\usepackage{amsmath}
\usepackage{amssymb}
\usepackage{latexsym}
\usepackage{subcaption}
\usepackage[colorlinks]{hyperref}
\usepackage{color}
\usepackage{enumerate}
\usepackage{enumitem}
\usepackage{cite}
\usepackage{makeidx}
\usepackage{colortbl}

\newcommand{\bra}{\begin{array}}
	\newcommand{\era}{\end{array}}
\newcommand{\beq}{\begin{equation}}
\newcommand{\eeq}{\end{equation}}
\newcommand{\beqar}{\begin{eqnarray}}
\newcommand{\eeqar}{\end{eqnarray}}

\def\BC{\bb C}
\def\_\BC{\bbi C}

\def\Tr {{\rm Tr}}

\def\( {\left(}
\def\) {\right)}
\def\[ {\left[}
\def\] {\right]}
\def\no2 {{\textstyle{n\over 2}}}

\def\Tr {{\rm Tr}}

\begin{document}
	\begin{titlepage}
		\setcounter{page}{1}
		\renewcommand{\thefootnote}{\fnsymbol{footnote}}

		\begin{flushright}
			%ucd-tpg 10-01\\
			%arXiv:yymm.xxxx
		\end{flushright}

\vspace{5mm}
\begin{center}
{\Large \bf {Time-Dependent Strain  in Graphene }}
\vspace{5mm}

{\bf Anha Bhat$^1$,     Salwa Alsaleh$^2$,  
Davood Momeni$^{3}$,   Atikur Rehman$^1$,  Zaid~Zaz$^4$, 
Mir~Faizal$^{5,6}$,  Ahmed Jellal$^{7}$, Lina Alasfar$^8$}

\vspace{5mm}

$^1${\em Department of Metallurgical and Materials Engineering,  National Institute of Technology, 
Srinagar, Kashmir-190006, India}

$^2${\em Department of Physics and Astronomy, College of Science, 
King Saud Universty, \\ Riyadh 11451, Saudi Arabia}
 
$^3${\em Department of  Physics,College of Science,Sultan Qaboos University\\
P.O.Box 36,P.C.123,Al-Khowd,Muscat,Sultanate of Oman}

$^4${\em Department of Electronics and Communication Engineering, University of Kashmir,  
Srinagar, Kashmir-190006, India}

$^5${\em Department of Physics and Astronomy, University of Lethbridge,  Lethbridge, AB T1K 3M4, Canada}

$^6${\em Irving K. Barber School of Arts and Sciences, University of British
Columbia - Okanagan,  
3333 University Way,  Kelowna,  British Columbia V1V 1V7, Canada}

$^7${\em Theoretical Physics Group, Faculty of Sciences, 
Chouaib Doukkali University,\\
PO Box 20, 24000 El Jadida, Morocco}

$^8${\em  Universit\'{e} Clermont Auvergne, \\
4, Avenue Blaise Pascal 63178 Aubi\`{e}re Cedex, France,}

\begin{abstract}
We will analyse the effect of time-dependent strain on a sheet of graphene by using the field theory approach.
It will be demonstrated that in the continuum limit,
such a strain will induce a non-abelian gauge field in graphene.
We will analyse the effective field theory 
of such  system near the Dirac points and study its topological properties.
\end{abstract}
\end{center}
\vspace{5mm}
\noindent PACS numbers:..............
\end{titlepage}

 \section{Introduction}
 
 Graphene is by far the most investigated  two-dimensional material both on  theroitical and experimental frontiers with wide applications which bridge condensed matter  to quantum mechanics when seen in the prespective of physics \cite{5b}
 because of its remarkable emergent  properties owning to different nanoscale phenomena which take place in its lattice. 
 One such aspect is the nanomechanics ,where the most important contribution is that of the  strain which is a  natural phenomena 
 that can  appear under different mechanism in graphene systems and therefore affecting its topology .
 %Experimentally, strains are expected to
 %arise naturally in graphene. 
 For example, graphene when grown on a 
 surface  usually experiences a moderate strain due to the
  surface corrugations of the substrate \cite{Teage} or the strain can aslo be developed when the  graphene lattice is not complimentary to the substrate on wheich it is grown\cite{Ni}. This can bring about the variations in the physical properties of the low dimensional systems and in this case the electronic properties are affected which again can pave a way to develop various applications .
  
 The  effective field theory describing graphene is a 
 massless fermionic theory in three dimensions \cite{5b}, which has also been verified experimentally  
\cite{6b,7b}. In such theory, the  deterministic  step is that the velocity of light $c$ is replaced by Fermi velocity $v_{\sf F} \approx \frac{c}{300}$.  
 It has been showed that a sheet of  graphene when under strain induce the effective  non-abelian gauge field \cite{nonnon,nonnon1}, which 
occur in the continum limit of the standard tight binding approach.  
This is because the spin connection of the graphene deformed metric can be related to the non-abelian gauge field. 
The strain in graphene sheets effects the Landau levels due to 
induced gauge fields \cite{laun} and  it can be comprehended  that these gauge fields have a role to play in  the physical 
behavior of graphene lattice . 
It has been reported that  the  time-dependent strain in graphene has been used to investigate  the topological electric current\cite{topo} and has also been used to 
obtain the fractional topological phases in its two dimensional arrangement\cite{topo1}. It is expected that the time-dependent strain will induce a three 
dimensional 
non-abelian gauge field theory in a deformed sheet of graphene, and we will construct such a theory in the present work. It may be noted that 
interesting deformations of  graphene have been studied using Dirac equation in curved spacetime \cite{8b}-\cite{10b}.  
Thus, it is possible to analyse graphene using the Dirac equation in curved spacetime, which produces an effective gauge field in it.

Motivated by different developments  on the subject cited above, we will analyse the gauge theory induced by a 
time-dependent strain in graphene. As the strain will be time-dependent, we shall consider the dynamics of the non-abelian gauge field and hence introduce its  kinetic term.  
It is possible to analyse geometry of curved spacetime  by using the translation group as a gauge group 
\cite{gauge11}-\cite{gauge22}
or
% there is a possibility  to study the curved spacetime by using 
Lorentz group as a gauge group \cite{gauge44}. These will be done by considering
 the metric of spacetime for the effective field theory describing the strained graphene
 and introducing a transformation in terms of the Lorentz group generators $SO(2,1)$.

%So, in this letter, we will analyze the geometry of a deformed sheet of graphene by using the formalism. 

The present paper is organized as follows. In section 2, we review some relevant tools
that include the tight-binding and the continuum models as a mathematical framework
to describe the graphene. In section 3,  we construct our effective theory describing
the strained graphene in terms of the non-abelian gauge field based on deformed spacetime.
The topological properties of the strained graphene will be done in section 4 where the instanton solutions will be obtained  numerically. Finally, we conclude our work and give some outlooks.

\section{Basic model}
Revisiting the basic information \cite{7b},
the  graphene sheet is a two-dimensional system made up of carbon atoms, which  
are arranged in an hexagonal honey comb structure. This arrangement  is a 
triangular lattice with a basis of two atoms per unit
cell where the lattice vector for this structure  are   
$
{\bf a}_1 = \frac{a}{2}\left(3, \sqrt{3}\right),  
{\bf a}_2 = \frac{a}{2}\left(3, -\sqrt{3}\right),$
and the carbon-carbon distance is $a \sim 1.42 \AA{}$. 
The reciprocal lattice vector for this structure 
 are  $
{\bf b}_1 = \frac{2\pi}{3a}\left(1, \sqrt{3}\right),  
{\bf b}_2 = \frac{2\pi}{3a}\left(1, -\sqrt{3}\right).$ 
The two points at the corners of graphene Brillouin zone are given by 
\begin{eqnarray}
K = \left(\frac{2\pi}{3a} , \frac{2\pi}{3\sqrt{3}a}\right), \qquad 
K'  = \left(\frac{2\pi}{3a} , -\frac{2\pi}{3\sqrt{3}a}\right) 
\end{eqnarray}
which are called the Dirac points.
The three nearest neighbor are located at    
\begin{eqnarray}
 \delta_1 =  \frac{a}{2}\left(1, \sqrt{3}\right), \qquad  \delta_2 = \frac{a}{2}\left(1, -\sqrt{3}\right), 
 \qquad \delta_3 = -a \left(1, 0\right)
\end{eqnarray}
and 
the six next to the nearest neighbors are located at 
\begin{eqnarray}
 \delta_1 =  \pm \bf{a}_1, \qquad \delta_2 = \pm \bf{a}_2, 
 \qquad \delta_3 = \pm \left(\bf{a}_1 - \bf{a}_2\right). 
\end{eqnarray}

 The electrons in the graphene 
can be described by the tight-binding approach where
the electrons can hop to both 
nearest and next to the nearest atoms. 
Now the Hamiltonian for this system can be represented by
 \begin{eqnarray}
  H = -t \sum (a^\dagger_\sigma b_\sigma + h.c)
  - t' \sum  ( a^\dagger_\sigma a_\sigma + b^\dagger_\sigma b_\sigma + 
  h.c.)
 \end{eqnarray}
where $\sigma$ denotes the spin,  $t\sim 2.8 eV$ 
is the hopping energy to the nearest neighbor, 
and $t'$ is the hopping energy to the next to the nearest 
neighbor. The energy bands for  this Hamiltonian 
can be written as \cite{tigh}
\begin{eqnarray}\label{energy}
 E_\pm = \pm t \sqrt{3  + f(k)} - t' f(k)
\end{eqnarray}
and  $f(k)$ is a function of the wave vector 
\begin{eqnarray}
 f(k)  = 2 \cos \sqrt{3} ( k_y a ) +
 4 \cos \left(\frac{\sqrt{3}}{2} k_y a \right) 
 \cos \left(\frac{3}{2} k_x a\right). 
\end{eqnarray}
Now expanding \eqref{energy} around the Dirac point $k = K + q$ (similarly for $K'$)  to obtain the linear dispersion relation
%it can be demonstrated that 
\cite{tigh}
\begin{eqnarray}
 E_\pm \sim  \pm v_{\sf F} |q|
\end{eqnarray}
where the Fermi velocity is $v_{\sf F}= \frac{3ta}{2}$.
 Here the velocity 
%$v_{f}\approx0.003 c$ is the Fermi velocity. 
%The Hamiltonian close to the Dirac point can be obtained.
 Expanding the operator in $\delta_i = \delta_1, \delta_2, \delta_3$, 
and using the approximation 
\begin{eqnarray}
 \sum_i \exp \left(\pm K \delta_i \right) = \sum_i \exp \left(\pm K' \delta_i \right) =0
\end{eqnarray}
%which is for Dirac point $K$
to obtain the Hamiltonian near it 
in the continuum limit making  
the Hamiltonian for this system
\beq
H_K= v_{\sf F} (\sigma_x p_x + \sigma_y p_y )
\eeq
which is
the usually denoted as Dirac Hamiltonian,  
with the velocity of light replaced by the Fermi velocity 
 \cite{tight}.
Now from this Hamiltonian, 
the effective field theory action describing graphene, in the continum limit,  
can be demonstrated as \cite{tight}
\begin{eqnarray}
S_{\sf eff} = \int d^3 x \bar \psi (i \gamma^\mu \partial_\mu )\psi 
= \int d^3 x \bar \psi i( \gamma^i \partial_i \psi 
- v_{\sf F} \gamma^0\partial_0 \psi )\psi.  
\end{eqnarray}

Having reviewed most of the fundamentals needed to describe the graphene system, it shall be in lieu of our interest to witness  as to how the  deformation in terms of strains in graphene  will effectuate the above results . More precisely, we  would construct the possible implication of 
a deformed spacetime  in a strained graphene.

\section{Strained graphene}
As reported , there are different experimental ways to realise the strain effect in graphene. 
%In this particular case ,the effective field theory of a strained graphene is represented by a three dimensional quantum field theory on
curved spacetime. The metric for the effective field theory of a sheet of  graphene under strain  
can be represented by \cite{9}
\begin{eqnarray}
 ds^2  = dt^2 - g^{ij}(x,y) dx^i dx^j. 
\end{eqnarray}
It may be noted that in $(2+1)d$-spacetime,  the Riemann tensor has only one independent 
component, which is proportional to the Gaussian curvature. So, we can represent the metric for strained graphene as 
\cite{10}
\begin{eqnarray}
 ds^2 = dt^2 - e^{2 \sigma (x,y)} dx^2 - e^{2 \sigma (x,y)} dy^2. 
\end{eqnarray}
So, the effect of strain on graphene is represented by a single function, 
$\sigma (x,y)$. 
However, it is also possible to take a time-dependent strain in graphene.
In fact, time-dependent strain in graphene has been used to analyse the 
a topological electric current in graphene \cite{topo} and %. It has also been 
%used 
also to obtain fractional topological phases in graphene \cite{topo1}. 
Now we can use the general consideration that the system does not contain terms which mix 
spatial and temporal coordinates, such as $dtdx$ and $dtdy$, and written a general time-dependent 
version of the metric for strain in graphene, without such terms. 
So, if    $\sigma (x,y,t)$ is a general function of time, then the metric describing graphene with time-dependent 
strain can be written as 
\begin{eqnarray}
 ds^2 = dt^2 - e^{2 \sigma (x,y, t)} dx^2 - e^{2 \sigma (x,y, t)} dy^2.
\end{eqnarray}
It may be noted that here we have not specified the form of  $\sigma (x,y,t)$, and kept it very general. 
Now the time can be corroborated as complex in terms of 
$\tau = it$ to obtain the new form for  the metric  
\begin{eqnarray}
 ds^2 = - d\tau^2 - e^{2 \sigma (x,y, \tau)} dx^2 - e^{2 \sigma (x,y, \tau)} dy^2. \label{metric}
\end{eqnarray}

To go further, we need to 
to introduce the following transformation 
$ x^a \longrightarrow x^b \Lambda^a_b$. 
This allows  a fermion in three dimension to transform as   
\begin{eqnarray}
 \psi  \longrightarrow U \psi = \exp \left(\frac{i}{2}\Lambda^{ab}(x) \Sigma_{ab} \right) \psi
\end{eqnarray}
where $\Sigma_{ab}$   describes  the generators of Lorentz group $SO(2,1)$, for $(t, x, y)$, and 
it  can be represented as  
$\Sigma_{ab} = -i [\gamma_a, \gamma_b]/4$. However, after complex equation, when  $(\tau, x, y)$, is used , the generators emerge as $SO(3)$.  
To obtain the geometry of a sheet of graphene using this transformation, 
 the metric is defined as 
$
 g_{\mu\nu} e^\mu_a e^\nu_b = \eta_{ab} 
$, and  $\omega_\mu^{ab}$ be  allowed to represent the spin connection associated with it. 
So, using the explicit expression for this spin connection $\omega_\mu^{ab}$ 
\cite{mart}, we could  define 
an effective non-abelian gauge field as 
\begin{eqnarray}\label{gfie}
 A_\mu = \omega_\mu^{ab} \Sigma_{ab}&=& [2 e^{\nu a}\partial_\mu e_\nu^b - 2 e^{\nu b} \partial_\mu e_\nu^a - 2 e^{\nu a}
 \partial_\mu e_\nu^b + 2 e^{\nu b}\partial_\nu e_\mu^a\nonumber \\ && + e_{\mu c}e^{\nu a} e^{\rho b}\partial_\rho e^c_\nu 
 - e_{\mu c} e^{\nu a} e^{\rho b} \partial_\nu e^c_\rho]\Sigma_{ab}. 
\end{eqnarray}
In line with above equations , the covariant derivatives of  effective non abelian guage field can be  elucidated  as                                                                                                                                    
\begin{eqnarray} 
 D_\mu =  \partial_\mu I + \frac{i}{2} A_\mu. 
\end{eqnarray}
 The gauge field \eqref{gfie} obtained from the spin connection transforms under $SO(3)$ as 
 \begin{eqnarray}
 A_\mu  %\omega_\mu^{ab}\Sigma_{ab} 
 \longrightarrow \left[ U  \omega_\mu U^{-1} -  (\partial_\mu U)  U^{-1}  \right]^{ab}
 = \left[ U A_\mu U^{-1} -  (\partial_\mu U)  U^{-1}  \right]^{ab}. 
 \end{eqnarray}
Now the effective field theory of graphene under strain can be written as 
\begin{eqnarray}
 S_{\sf eff } &=&  \int d^3 x e \bar \psi  \left(i \gamma^a e_a^\mu D_\mu  \right) \psi\nonumber \\
 &=&  \int d^3 x e \bar \psi  \left(i \gamma^a e_a^\mu \left( \partial_\mu I + \frac{i}{2} A_\mu \right)\right) \psi\nonumber \\
  &=&  \int d^3 x e^{2\sigma \left( x,y,\tau \right)}  \bar \psi  \left(i \gamma^a e_a^\mu \left( \partial_\mu I + \frac{i}{2}  
  \omega_\mu^{ab} \Sigma_{ab}\right) \right) \psi
\end{eqnarray}
which  is invariant under the $SO(2, 1)$ gauge symmetry. 
It is possible for the strain 
to fluctuate in time . In order to analyse its effect the  kinetic part of this 
effective non-abelian gauge theory has to be calculated keeping in view that the detailed study of electronic and mechanical properties of graphene have been already accomplished .\cite{strain1,strain2}. 
In fact, it has been proposed that  the strain can be used to  generate large effective gauge fields, which 
can have direct consequences on the electronic  properties of graphene \cite{strain6,strain7}. 
which can be harnessed to develop  quantum electronic pump based applications based on this proposal \cite{pump,pump80}.

An important property for the above strain is that the covariant derivatives 
do not commute when  present in a sheet of graphene. In fact, 
the commutator of these two covariant derivatives can be expressed as 
\begin{eqnarray}
 iF_{\mu\nu}^{ab}\Sigma_{ab}=
 [D_\mu, D_\nu] = T^c_{\mu \nu}D_c + \frac{i}{2}R^{ab}_{\mu\nu} \Sigma_{ab}.
\end{eqnarray}
If it is  assumed that  there is no contribution pertaining to the torsion
$T^c_{\mu \nu}$ term,  $T^c_{\mu\nu} =0$, then the commutator of two covariant derivatives for graphene can be expressed using the
   Riemann tensor as 
\begin{eqnarray} iF_{\mu\nu}^{ab}\Sigma_{ab} &=& R^{ab}_{\mu\nu}\Sigma_{ab} \nonumber\\
&=& \partial_\mu A_{\nu} - \partial_\nu A_\mu + [A_\mu, A_\nu].
\label{r}
\end{eqnarray}
The curvature tensor  transforms as 
\begin{eqnarray}
  iF_{\mu\nu}^{ab} &=&  R^{ab}_{\mu\nu} \to [U]^a_c R_{\mu\nu}^{cd} [U^{-1}]^{b  }_{d } \nonumber \\ &=&  i [U]^a_c F_{\mu\nu}^{cd} [U^{-1}]^{b  }_{d }. 
\end{eqnarray}
As the strain is time dependent, it changes with time, which necessitates  the introduction of 
a Kinetic term  for the gauge field produced by  strain.  
The simplest action, that can produce dynamics, without higher derivative ghost states, is the 
Yang-Mills action. Then, the Kinetic action for the gauge fields produced by strain 
can be written as 
\begin{eqnarray}
S_g&=& - \frac{1}{4g^2 } \int d^3 x e \,  Tr (F^{\mu\nu} F_{\mu\nu})  \\ \nonumber 
&= &   \frac{1}{4g^2 } \int d^3 x  e^{2\sigma \left( x,y,\tau \right)} \left[\, 2\left( {\frac {\partial ^{2}}{\partial {\tau}^{2}}}\sigma \left( x
,y,\tau \right) + \left( {\frac {\partial }{\partial \tau}}\sigma
 \left( x,y,\tau \right)  \right) ^{2} \right) ^{2}\right. \\&&\nonumber+{\frac { \left( {
\frac {\partial ^{2}}{\partial {y}^{2}}}\sigma \left( x,y,\tau
 \right) +{\frac {\partial ^{2}}{\partial {x}^{2}}}\sigma \left( x,y,
\tau \right) + \left( {\frac {\partial }{\partial \tau}}\sigma \left( 
x,y,\tau \right)  \right) ^{2}{{\rm e}^{2\,\sigma \left( x,y,\tau
 \right) }} \right) ^{2}}{ \left( {{\rm e}^{2\,\sigma \left( x,y,\tau
 \right) }} \right) ^{2}}}\\ && \nonumber \left. -{{\rm e}^{2\,\sigma \left( x,y,\tau
 \right) }} \left( {\frac {\partial ^{2}}{\partial y\partial \tau}}
\sigma \left( x,y,\tau \right)  \right) ^{2} -{{\rm e}^{2\,\sigma
 \left( x,y,\tau \right) }} \left( {\frac {\partial ^{2}}{\partial x
\partial \tau}}\sigma \left( x,y,\tau \right)  \right) ^{2}\right],  
\label{sg}
\end{eqnarray}
where $g$ is a suitable constant describing this theory, where   %This constant $g$
physically denotes the strength  by coupling the fluctuations in the geometry to the 
effective field theory describing fermions in graphene 
under strain.
Now the effective field theory for system can be stated as  
\begin{eqnarray}
 S_{t} = S_{\sf eff}  + S_g 
\end{eqnarray}
 which is the effective field theory of a sheet of graphene under time dependent strain, 
 such that the  dynamics is associated with the strain.

 \section{Topological properties}

 In view of the above effective 
 field theory ,there is a possibility to study  its implications regarding the topological properties of graphene.
 It may be noted that topological defects have been observed in graphene \cite{td}-\cite{td12}.
 In fact the bosonic symmetry protected topological 
 state has been used to 
 study the transport in graphene. 
 The instanton tunneling is important to understand the shot noise measurement for such topological state in the strong interaction limit \cite{stin} with respect to the instanton solution 
and the in-gap fluctuation states in biased bilayer graphene \cite{inst}. Therefore, it will be interesting to understand the behavior of the
 instanton solution of the action $S_g$. 
 We can complexify time, and with such  
Euclidean time, our theory can be analyzed as a theory on a three dimensional
manifold $M$ with Euclidean signature with a   
discussion about  the topological properties of the instanton 
for this manifold.

Since the action 
 is in fact an Euclidean action, the instanton solution is divided into two components,
 the self-dual and anti self-dual solutions.
 
 \begin{eqnarray}
 F_{\mu\nu} = \pm *F_{\mu \nu}
 \end{eqnarray}
 with $*$ denoting the hodge star operator. Using the explicit expression for $F_{\mu \nu}$ from~\eqref{r}, 
  it is found to have three independent components:
\begin{eqnarray}
F_{\tau x} &=& \frac{1}{2}\left( e^{2 \sigma} (\partial_\tau \sigma)^2 + \partial^2_y \sigma +\partial^2_x \sigma \right) [\gamma_\tau,\gamma_x]  
\\&&\nonumber+ \frac{1}{2} (\partial_\tau \partial_y \sigma) [ \gamma_\tau, \gamma_y]+ \frac{1}{2} (\partial_\tau \partial_x \sigma) [ \gamma_x, \gamma_y]\\
F_{\tau y} &=& \frac{1}{2}(\partial_\tau \partial_y \sigma) [ \gamma_\tau, \gamma_x]+\frac{1}{2}\left(  (\partial_\tau \sigma)^2+ \partial^2_\tau \sigma\right) 
[ \gamma_\tau, \gamma_y] \\
F_{xy} &=& \frac{1}{2} (\partial_\tau \partial_x \sigma ) [ \gamma_\tau, \gamma_x] + \frac{1}{2}\left( e^{2 \sigma}
( (\partial_\tau \sigma)^2+\partial^2_\tau \sigma)\right) [ \gamma_x, \gamma_y].
\end{eqnarray}
The self-dual (instanton) solution,  yields  the following differential equations
\begin{eqnarray}
&& e^{2 \sigma} (\partial_\tau \sigma)^2) + \partial^2_y \sigma +\partial^2_x \sigma =
- (\partial_\tau \partial_y \sigma)-(\partial_\tau \partial_x \sigma) \label{pde1} \\
&& (\partial_\tau \partial_x \sigma) = -  e^{2 \sigma}( (\partial_\tau \sigma)^2+\partial^2_\tau \sigma) \label{pde2}\\
&& (\partial_\tau \partial_y \sigma) =  (\partial_\tau \sigma)^2-\partial^2_\tau \sigma \label{pde3}.
\end{eqnarray}
Substituting the LHS of \eqref{pde2} and \eqref{pde3} in \eqref{pde1}, to end up with
the partial differential equation for the strain function $ \sigma(\tau, x,y)$ for the instanton
\begin{eqnarray}
(1+ e^{2 \sigma}) \partial^2_\tau \sigma +(\partial_\tau \sigma)^2  + \partial^2_x \sigma + \partial^2_y \sigma = 0 
\end{eqnarray}
as well as 
the differential equation 
for the anti-instanton solution
\begin{eqnarray}
(1- e^{2 \sigma}) \partial^2_\tau \sigma -(\partial_\tau \sigma)^2  + \partial^2_x \sigma + \partial^2_y \sigma = 0.
\end{eqnarray}
These PDE's can be solved for different boundary conditions, depending on the characteristics of the graphene 
under consideration.   For example, 
if the Dirichlet boundary conditions are introduced on the $3D$ Euclidean plane, 
letting $ \sigma(\tau =0 ,x,y)=0$ and $ \partial_\tau \sigma(\tau=0,x,y)=0$, the instanton and anti-instanton solutions, near $ \tau \approx 0$,
become
\begin{eqnarray}
&& 2\partial^2_\tau \sigma +  \partial^2_x \sigma + \partial^2_y \sigma = 0\\
&&
\partial^2_\tau \sigma +  \partial^2_x \sigma + \partial^2_y \sigma = 0
\end{eqnarray}
which can be solved numerically, see Figure~1:

 \begin{figure}[htp!]\label{inst}
 	\centering
 	\begin{minipage}[b]{0.44\textwidth}
 		\includegraphics[width=\linewidth]{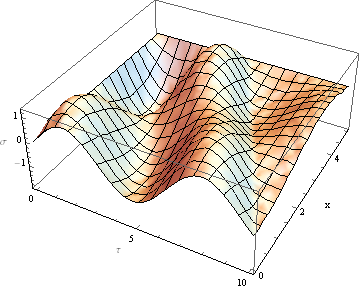}
 	\end{minipage}
 	\hfill
 	\begin{minipage}[b]{0.44\textwidth}
 		\includegraphics[width=\linewidth]{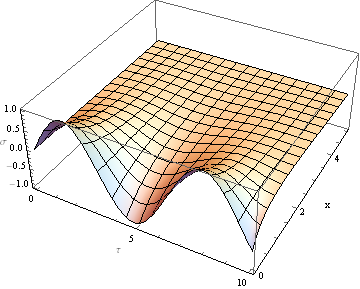}
 	\end{minipage}
 	 		\label{inst}
 		\caption{Plot of the instanton (left) and anti-instanton (right) solutions for $ \sigma (\tau,x)$, near $ \tau=0$.
 		Here we reduced  the dimensions by suppressing the $y$ dependence, and plotted the behavior 
 		of the function strain function, with respect to $x$ and $\tau$.   }
 \end{figure}

 We can complexify time, and with such an 
Euclidean time, this theory can be analyzed as a theory on a three dimensional
manifold $M$ with Euclidean signature, such as
\begin{eqnarray}
S_g^{E} = \mp \frac{1}{2g^2} \int_M \Tr\left( \mathbf{F} \wedge \mathbf{F} \right).
\label{se}
\end{eqnarray}
We can now  discuss the topological properties of the instanton 
for this manifold. It is possible to compactify $M$ 
 into a closed surface,
$M'$ by assuming a characteristic length $L$ and periodic boundary conditions, such that $ x \longrightarrow L$~\cite{nak}.
Now for this compact manifold, we have 
\begin{eqnarray}
A_\mu = U(x)^{-1} \partial_\mu U(x)
\end{eqnarray}
where $ U(x)$ are elements of the $SO(3)$ gauge symmetry. 
By this compactification and Poincare lemma, we can write $F^2$ as an exact form
\begin{eqnarray}
\Tr \mathbf{F}^2 = d\mathbf{K}
\end{eqnarray}
and $\mathbf{K}$ is a local three form, which implies that 
$\Tr \mathbf{F}^2$ is an element of the cohomology group $ H^3(M')$. 
Moreover, it is possible to write $\mathbf{K}$ as  a Chern Simons form~\cite{cs}
\begin{eqnarray}
\mathbf{K} = \Tr \left(\mathbf{A}\, d\mathbf{A}+ \frac{2}{3}\mathbf A^2\right).
\end{eqnarray}
Now by using the  Stokes theorem, we can write the action~\eqref{se} after compactification as 
\begin{eqnarray}
S^E_g =- \frac{1}{4 g^2} \int_{\partial M'} \Tr\left( \mathbf A^3\right).
\end{eqnarray}
For the instanton solution, we have the expression for the gauge field  ${\bf A}$
in terms of the mappings $ U: S^2 \longrightarrow SO(2,1)$. Then the above integral 
is reduced to the degree of that mapping divided by the area of the unit   $S^2$
\begin{eqnarray}
\frac{1}{6 \pi g^2}\int_{\partial M'} \Tr\left(\mathbf  A^3\right)= \frac{n}{g^2}
\end{eqnarray}
which can be rewritten  as
\begin{eqnarray}
\frac{1}{2}\int_{M'} \Tr\left( \frac{i\mathbf F}{2 \pi}\right) ^2.
\end{eqnarray}
By using the definition of the field strength tensor in terms of the Riemann curvature given 
in \eqref{r},
the above integral reduces to the Euler characteristic of the 
compactified manifold $M'$~\cite{Fujiwara:2012vk,Cano:2017sqy}
\begin{eqnarray}
\frac{1}{2}\int_{M'} \Tr\left( \frac{i\mathbf F}{2 \pi}\right) ^2 
= \frac{1}{8 \pi^2}\int_{M'} \Tr\left( \mathbf R \wedge \mathbf R\right) = \chi(M').
\end{eqnarray}  
Hence, we observe that the degree of the mapping corresponding to instanton solution  
is proportional to a topological invariant of the theory. Therefore, it is possible to classify corrugated  sheet
of graphene under dynamic time dependent strain by 
using these topological instantons. It clear that,  these topological  instantons 
can produce measurable effects.

\section{Conclusion}

 Using the field theory approach, a sheet of graphene was analysed  with a time-dependent strain. 
 The spin connection of this theory was used to obtain a non-abelian
gauge field. As the strain was time-dependent, the 
kinetic term for this  
effective non-abelian gauge theory was also considered  which motivated the  study of the topological properties of the system of interest given its deformations and shows that
the topological  instantons  could map the considerable effects  as the result of such a proposed study.
There are various interesting deformations of the effective field theory of graphene, and these can be analysed using various interesting and equally 
effective geometries. 
The  surface of revolution with constant negative
curvature  has  been constructed  using graphene \cite{1a}. We investigated 
the instanton solution for the effective Yang-Mills theory, 
and analyzed  the Euler characteristic of such a sheet of  graphene. 

It is  possible to analyse  the analogous  black hole  like solution   in  the effective field theory describing
 graphene \cite{1}, using a BTZ  (Banados Teitelboim Zanelli) \cite{2a}-\cite{BTZ} like metric. 
 It may be noted that this is only an effective solution 
and in which  the velocity of light is replaced by the Fermi velocity. Thus, just as a real black hole forms the horizon 
for  light, these analogous black hole like solutions form effective horizons for Fermi velocity.   
It has been demonstrated that a  deformed sheet of  graphene  can also be analysed using negative constant curvature in an externally
applied magnetic field, which  can be done by using    
 a stationary optical metric of the Zermelo
form that is conformal to the BTZ  black hole \cite{1}.  
Thus, interesting  analogous geometries have been studied using the effective field theory of graphene. 
Therefore, an analogous black hole like solution can form in a deformed sheet of  graphene.
However, it is known that the usual black hole physics leads to the existence of the generalized uncertainty principle \cite{z4}-\cite{z5}. 
It would be interesting to analyse this relation between analogous black holes in graphene and generalized uncertainty principle which would further develop investigations at the crossroads of physics pertaining to photonics and optoelectronic applications.

\section*{Acknowledgments}
The authors  would like to thank Maria A. H. Vozmediano, 
and  Alfredo Iorio for  their useful comments.

  \end{document}